\documentclass[11pt]{article}
\usepackage{color}
\usepackage{amssymb}
\usepackage{amsmath}
\usepackage{epsfig}

\definecolor{pink}{rgb}{1,.4,.7}
\definecolor{magenta}{rgb}{1,0,1}
\definecolor{violet}{rgb}{.9,.25,.6}
\definecolor{darkolivegreen3}{rgb}{.6,.8,.35}
\definecolor{maroon3}{rgb}{.8,.26,.56}
\definecolor{mediumorchid}{rgb}{.73,.33,.83}
\definecolor{mediumorchid1}{rgb}{1.,.33,.63}
\definecolor{darkgreen}{rgb}{0.1,.6,.13}
\definecolor{lightyellow}{rgb}{1.,1.,.82}
\definecolor{turquoise}{rgb}{.35,.80,.71}
\definecolor{coral}{rgb}{1.,.6,.21}
\definecolor{orangered}{rgb}{1.,.5,0.}
\definecolor{orange}{rgb}{1.,.65,.1}
\definecolor{blue1}{rgb}{.48,.53,1.}
\definecolor{gold}{rgb}{1.,.85,0.}
\definecolor{darkviolet}{rgb}{.54,.04,.84}

\def\Journal#1#2#3#4{{#1} {\bf #2}, (#3) #4}

\def\etal{{\it et al.}}
\def\AA{\em A.\& A.}

\def\APJ{\em ApJ.}

\def\IET{\em IEEE Trans. Comm.}

\def\MRA{\em MNRAS}

\def\NPB{{\em Nucl. Phys.} B}

\def\PLB{{\em Phys. Lett.}  B}

\def\PRD{{\em Phys. Rev.} D}

\def\filepath{./} 

\def\be{\begin{equation}}
\def\ee{\end{equation}}
\def\bea{\begin{eqnarray}}
\def\eea{\end{eqnarray}}

\begin{document}
\begin{center}
\Large \bf {Nonparametric determination of the sign of w+1 in the equation of state of Dark Energy}\\
\end{center}

\begin{center}
{\it Houri Ziaeepour\\
{Mullard Space Science Laboratory,\\Holmbury St. Mary, Dorking, Surrey
RH5 6NT, UK.\\
Email: {\tt hz@mssl.ucl.ac.uk}}
}
\end{center}

\medskip
\medskip

\begin {abstract}
We present a nonparametric method to determine the sign of $w+1$ in the 
equation of state of dark energy. It is based on geometrical behaviour and is 
more tolerant to uncertainties of 
other cosmological parameters than fitting methods. It permits to distinguish 
between different classes of dark energy models even with relatively low 
precision data. We apply this method to SNLS supernovae and to gold sample of 
re-analyzed supernovae data from Riess \etal 2004~\cite{sncp}. Both data sets 
show strong indication of $w < -1$. If this result is confirmed by more 
extended and precise data available in near future, many of dark energy models, 
including simple cosmological constant, standard quintessence models without 
interaction between quintessence scalar field(s) and matter, and scaling 
models are ruled out.
\end {abstract}

Recent observations of SuperNovae (SN), Cosmic Microwave Background (CMB), and 
Large Scale Structures (LSS) indicate that the dominant content of the 
Universe is a mysterious energy with an equation of state very close to 
Einstein Cosmological Constant. The equation of state is defined by $w$, 
the ratio of pressure $p$ to density $\rho$, $w = P/\rho$. For a cosmological 
constant $w = -1$. The observed mean value of $w$ for dark energy is very 
close to $-1$. Some of the most recent estimations are the followings: From 
combination of 3-year WMAP and SuperNova Legacy Survey (SNLS), 
$w = -0.97^{+0.07}_{0.09}$~\cite{wmapeyear}; from combination of 3-year WMAP, 
large scale structure and supernova data, 
$w = -1.06^{+0.016}_{-0.009}$~\cite{wmapeyear}; from combination of CMAGIC 
supernovae analysis and baryon acoustic peak in SDSS galaxy clustering 
statistics at $z = 0.35$, $w = -1.21^{+0.15}_{-0.12}$~\cite{cmagicsn}; 
and finally from baryon acoustic peak alone $w = -0.8 \pm 0.18$. It is evident 
that with inclusion of one or two sigma uncertainty to measured mean values, 
the range of possible values for $w$ runs across the critical value of 
$-1$. Moreover, in all these measurements the value of $w$ depends on 
other cosmological parameters and their uncertainties in a complex way.
Reconstruction methods for determining cosmological parameters from 
observations (see ~\cite{reconsrev} and references therein for a review of 
methods) usually use fitting of continuous parameters on the data and 
determine a range for $w$. 

On the other hand, the sign of $\gamma \equiv w+1$ is more crucial for 
distinguishing between various dark energy models than its exact value. For 
instance, if $\gamma < 0$, scalar field (quintessence) models with 
conventional kinetic energy and potential are ruled out because in these 
models $\gamma$ is always positive. Decay of dark matter to dark 
energy~\cite {houriustate}~\cite {houridmquin}, or in general an interaction 
between these components can lead to an effective $\gamma < 0$ without 
violating null energy condition~\cite {quinint}~\cite {houridmquin1}. 

Here we propose a nonparametric method specially suitable for 
estimating the sign of $\gamma$. When the quality of data is adequate, the 
quantity $A(z)$ defined in (\ref{densderiv}) can also be used to fit the data 
and to measure the value of $\gamma$. The expression {\it nonparametric} here 
is borrowed from signal processing literature where it means testing a null 
hypothesis against an alternative hypothesis by using a discrete condition 
such as jump, sign changing, etc., in contrast to constraining a continuous 
parameter (see e.g. ~\cite{nonparam}). We show that geometrical 
properties of $A(z)$ are related to the sign of $\gamma$ and we can detect 
it without fitting a continuous parameter.

The density of the Universe at redshift $z$ is:
\be
\frac {\rho (z)}{\rho_0} = \Omega_m (1+z)^3 + \Omega_h (1+z)^4 + 
\Omega_{de} (1+z)^{3\gamma} 
\label{density}
\ee
where $\rho (z)$ and $\rho_0$ are total density at redshift $z$ and in 
local Universe, respectively; $\Omega_m$, $\Omega_h$, and $\Omega_{de}$ 
are respectively cold and hot matter, and dark energy fraction in the 
total density at $z=0$. We consider a flat universe in accordance with 
recent observations~\cite{wmapeyear}. At low redshifts, the contribution 
of CMB to the total mass of the Universe is negligible. The contribution 
of neutrinos is ${\Omega}_{\nu} h^2 = \sum m_{\nu} / 92.8~eV$, 
$h \equiv H_0 / 100 km~Mpc^{-1}sec^{-1}$, where $H_0$ is present Hubble 
constant. The upper limit on the sum of masses of neutrinos from 
3-year WMAP is $\sum m_{\nu} < 0.62$ ($95\%$ confidence 
level)~\cite{omeganu}. Therefore, for $z < 1$ their contribution to the 
total mass of the Universe is $\lesssim 4\%$ even if one of the neutrinos 
has very small mass and behaves as a warm dark matter. This is less than 
the uncertainty on the fraction of dark matter, and thus the approximation 
$\Omega_m + \Omega_{de} \approx 1$ is justified. It can be easily shown that 
in this case:
\be
{\mathcal A}(z) \equiv \frac{1}{3 (1+z)^2 \rho_0} \frac {d\rho}{dz} - 
\Omega_m = \gamma \Omega_{de}(1+z)^{3 (\gamma - 1)} \label {densderiv}
\ee
Similar expressions can be obtained for non-standard cosmologies such as 
DGP~\cite{dgp} model and other string/brane inspired 
cosmologies~\cite{branecosmos}. It is also possible to find an expression 
similar to (\ref{densderiv}) for non-flat FLRW models and without neglecting 
hot matter. The left hand side would however depend on $\Omega_k$, and 
$\Omega_h$ and would be more complex. Nonetheless, when the contribution of 
these components at low redshifts are much smaller than cold matter and dark 
energy, the general behaviour of ${\mathcal A}(z)$ will be the same as 
approximate case studied here.

\begin{figure}[ht]
\begin{tabular}{cc}
{\bf a} \epsfig{figure=\filepath/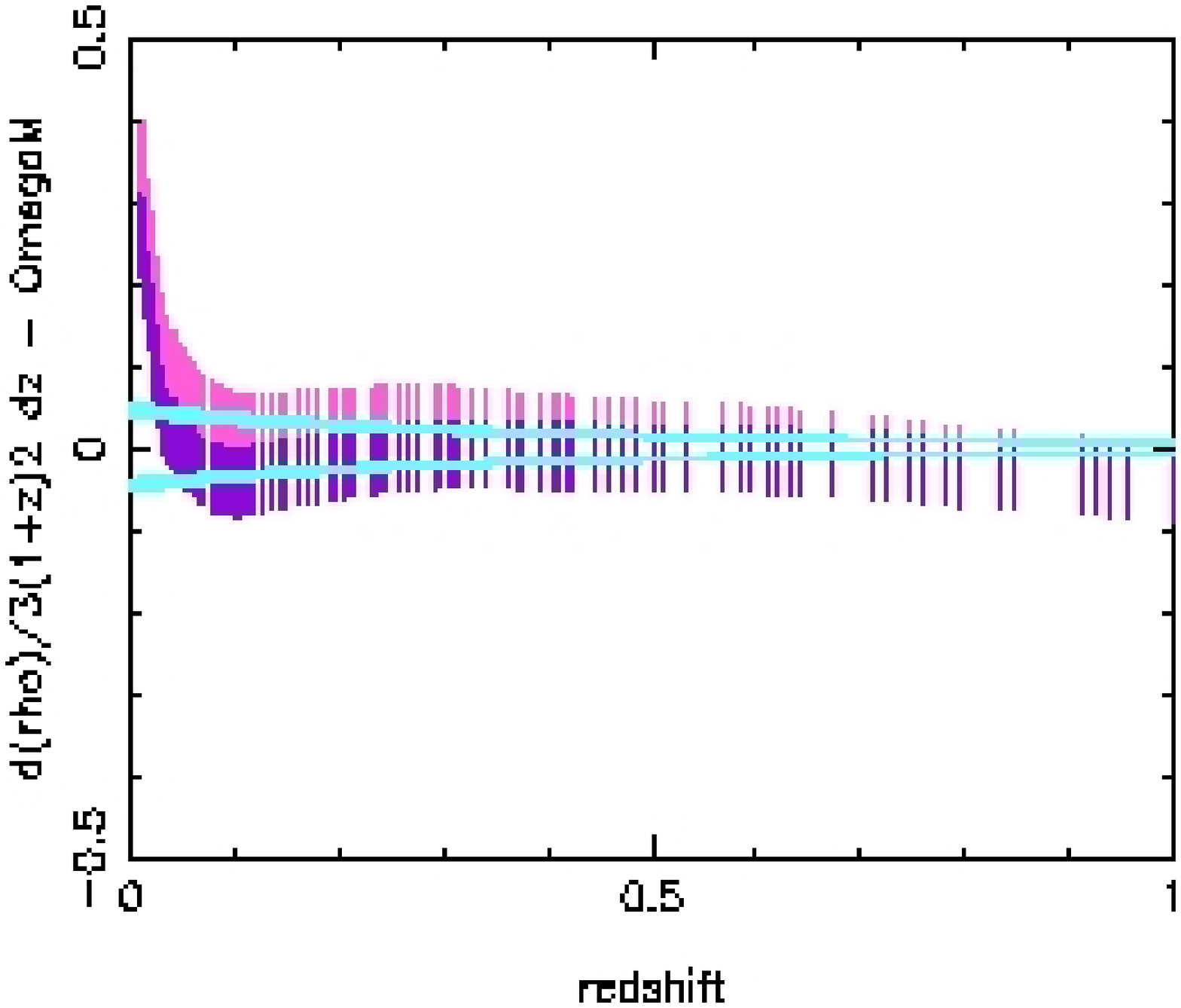,width=7cm} &
{\bf b} \epsfig{figure=\filepath/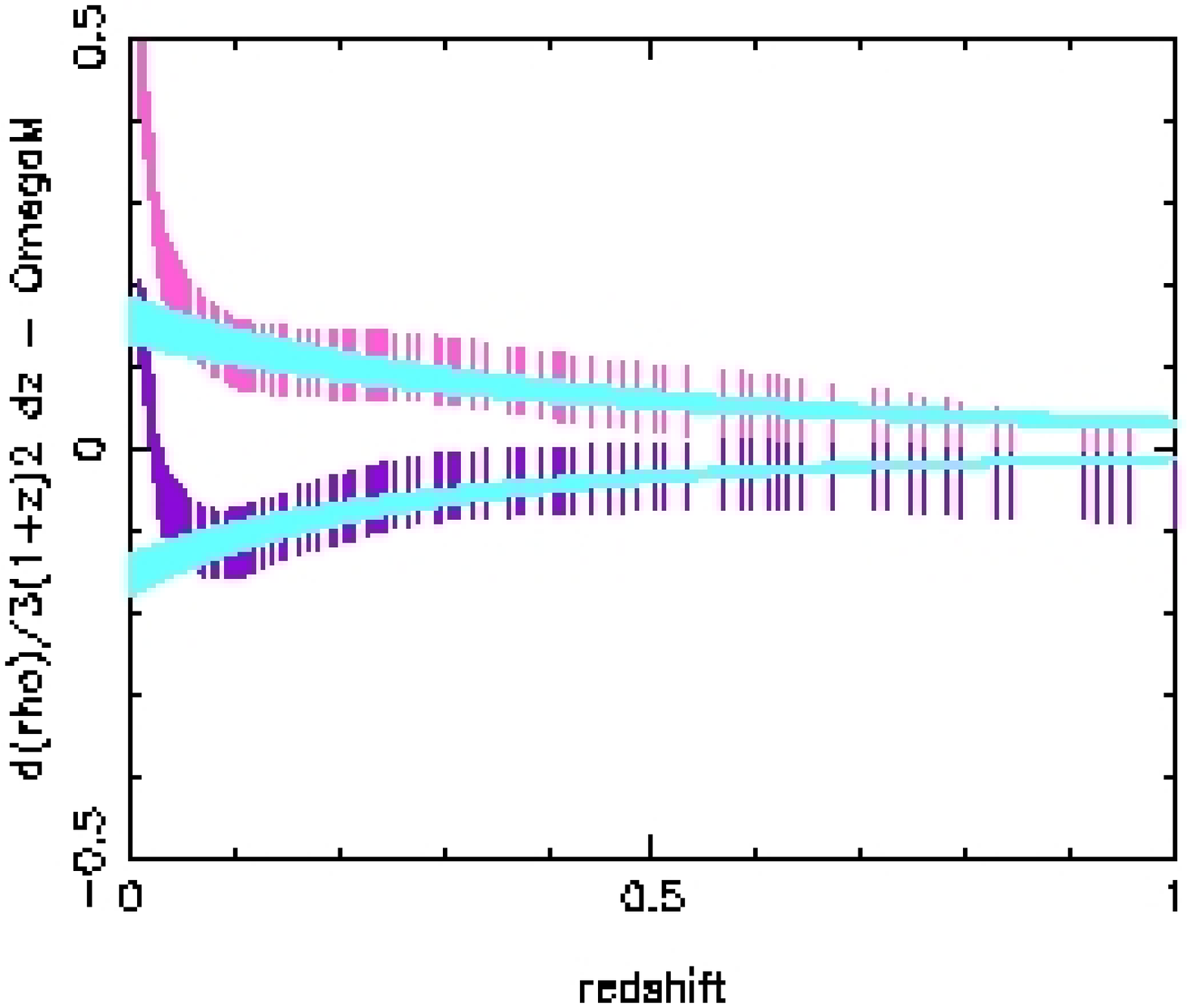,width=6.5cm}
\end{tabular}
\caption{${\mathcal A}(z)$ in Eq.(\ref{densderiv}) as a function of 
redshift; {\bf a}: $\gamma = \pm 0.06$ (corresponding to best fit of 3-year 
WMAP, LSS, and SN data~\cite{wmapeyear}) and {\bf b}: $\gamma = \pm 0.2$ 
(corresponding to best fit of 3-year WMAP and CMAGIC SN data~\cite{cmagicsn}). 
For both plots $H_0 = 73~km~Mpc^{-1}~sec^{-1}$ and $\Omega_{de} = 0.77$. We 
consider $5\%$ of uncertainty for both quantities. 
Blue curves present right hand side of (\ref{densderiv}), including the 
uncertainty on the value of $\Omega_{de}$. Curves decreasing or increasing 
with increasing redshift present respectively positive and negative 
$\gamma$. Magenta and purple curves present numerical determination of 
${\mathcal A}(z)$ respectively for positive and negative $\gamma$ from about 
150 simulated supernovae distributed with uniform probability in $\log (z)$. 
These plots show that numerical errors dominate near two boundaries of the 
redshift range. It is also evident that with this method one can distinguish 
between positive and negative $\gamma$ for $|\gamma| \gtrsim 0.06$. 
\label {fig:theorycurve}}
\end{figure}

The right hand side of (\ref{densderiv}) has the same sign as $\gamma$. 
Moreover, the sign of its derivative is opposite to the sign of $\gamma$ 
because due to the smallness of observed $\gamma$, the term $\gamma -1$ is 
negative. This means that ${\mathcal A}(z)$ is a concave or convex function 
of redshift, respectively for positive or negative $\gamma$, see figure 
\ref{fig:theorycurve}. In the case of a cosmological constant 
${\mathcal A}(z) = 0$ for all redshifts. This second feature of expression 
(\ref{densderiv}) is interesting because if $\Omega_m$ is not correctly 
estimated, ${\mathcal A}(z)$ will be shifted by a constant, but this will not 
modify geometrical properties of ${\mathcal A}(z)$.

The left side of expression (\ref{densderiv}) can be directly estimated 
from observations. More specifically, $\Omega_m$ is determined from 
conjunction of CMB, LSS, and supernova type Ia data, and at present it 
is believed to be known with a precision of $\sim 5\%$. At low redshifts, 
the derivative of the density is best estimated from SN type Ia observations. 
In the case of FLRW cosmologies, the density and its derivative can be 
related to luminosity distance $D_l$, and its first and second derivatives:
\bea
&& {\mathcal B}(z) \equiv \frac{1}{3 (1+z)^2 \rho_0} \frac {d\rho}{dz} = 
\frac{\frac{2}{1+z}(\frac{dD_l}{dz} - \frac {D_l}{1+z}) - 
\frac{d^2D_l}{dz^2}}{\frac{3}{2} (\frac{dD_l}{dz} - \frac {D_l}{1+z})^3}
\label {rhoderdl} \\
&& D_l = (1+z) H_0 \int_0^z \frac {dz}{H (z)} \quad , \quad 
H^2 (z) = \frac{8\pi G}{3} \rho (z) \label {dlh}
\eea
It is remarkable that the right hand side of (\ref{rhoderdl}) depends only 
on one cosmological parameter, $H_0$. Nonetheless, similar to an uncertainty 
on $\Omega_m$, $H_0$ scales ${\mathcal B}(z)$ similarly at all redshifts, 
and therefore it 
does not change the overall geometrical properties ${\mathcal B}(z)$ and 
${\mathcal A}(z)$. $D_l$ can be directly obtained from observed luminosity of 
standard candles such as supernovae type Ia. In the case of LSS observations 
where the measured quantity is the evolution of density $\rho$ with redshift, 
${\mathcal B}(z)$ is measured directly up to an overall scaling by 
$\rho_0$. This does not change the geometrical properties of 
${\mathcal B}(z)$ and ${\mathcal A}(z)$, i.e. the scaling by a positive 
constant does not flip a convex curve to concave or vis-versa. 

In summary, uncertainties of $\Omega_m$ and $H_0$ do not affect the detection 
of the sign of $\gamma$ through geometrical properties of ${\mathcal A}(z)$. 
This is quite different from fitting methods. They are sensitive to all 
numerical parameters $H_0$, $\Omega_m$, $\Omega_{de}$, and $w$ in a complex way, 
usually through a non-linear equation such as chi-square or likelihood 
equation, and it is very difficult to assess the effect of uncertainty of one 
parameter on the estimation of others, and more specifically on the 
determination of the sign of $\gamma$.

When this sign detection method is applied to standard candle data such as 
supernovae type Ia where the measured values are related to $D_l$, according 
to (\ref{rhoderdl}), one has to calculate first and second derivative of $D_l$.
Numerical calculation of derivatives is not trivial. To have a stable and 
enough precise result, not only the data must have high resolution and low 
scatter, but also it is necessary to smooth them. To test the stability of 
numerical calculation and the method in general, we also apply it to simulated 
data. The details of numerical methods is discussed in the Appendix. 

\begin{figure}[ht]
\begin{tabular}{cc}
{\bf a} \epsfig{figure=\filepath/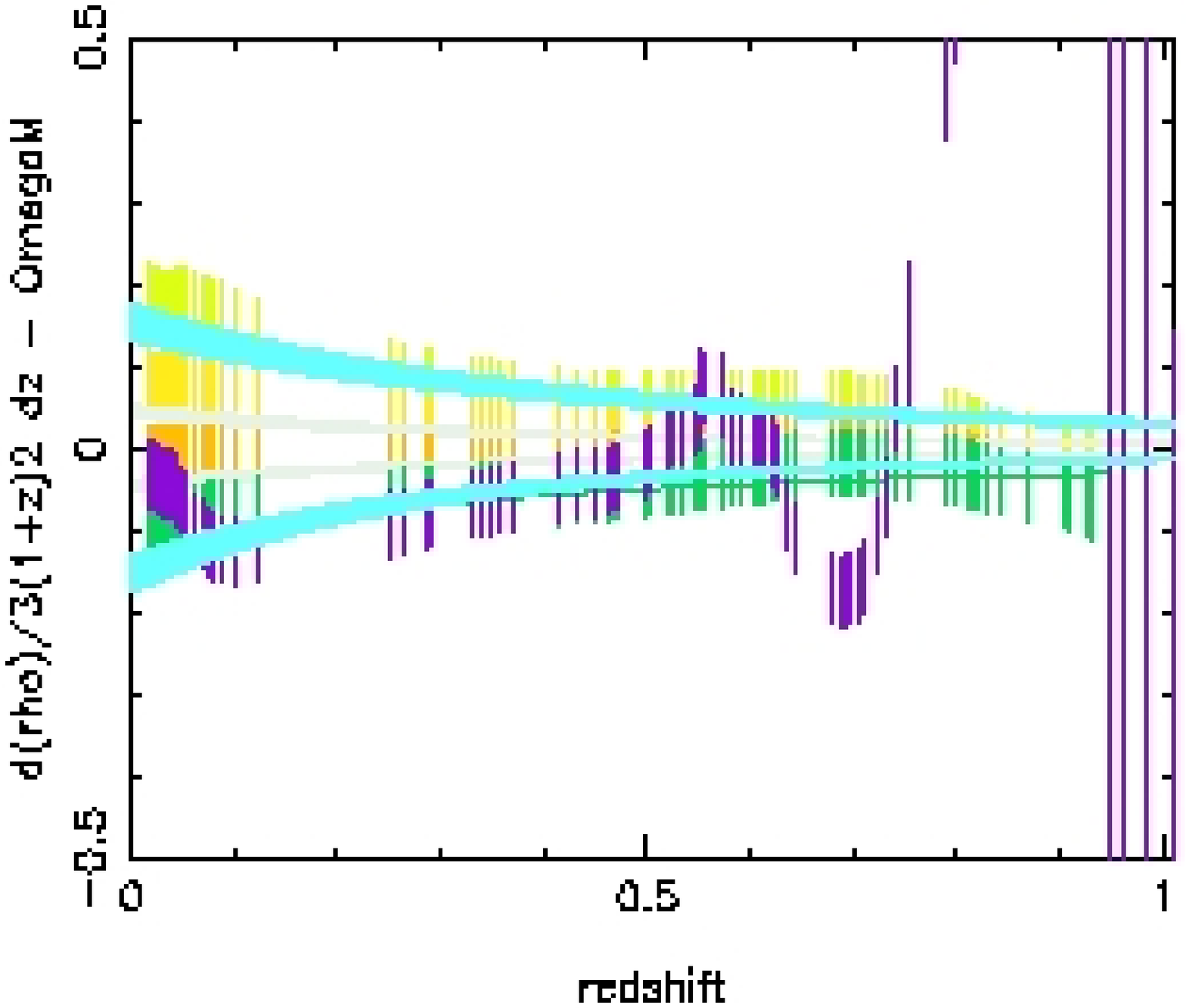,width=6.5cm} &
{\bf b} \epsfig{figure=\filepath/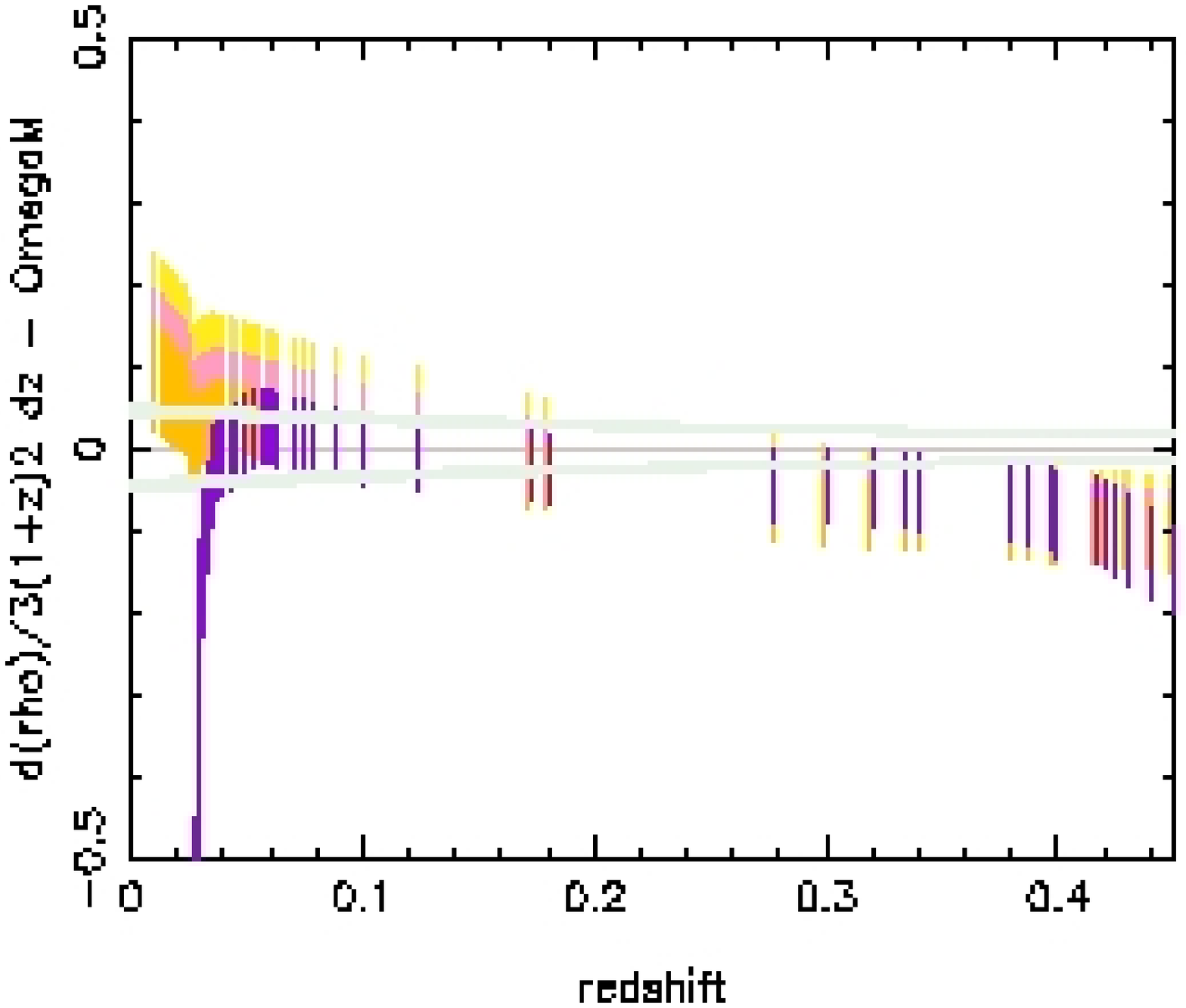,width=6.5cm}
\end{tabular}
\caption{${\mathcal A}(z)$ from: 117 supernovae of the SNLS data 
({\bf a}-purple curve), and 88 supernovae with $z < 0.45$ recompiled and 
re-analyzed by Riess \etal 2004~\cite{sncp} ({\bf b}-purple curve). Error 
bars present 1-sigma uncertainty. In ({\bf a}), green, orange, yellow, and 
light green curves present the reconstruction of ${\mathcal A}(z)$ from 
simulations for $\gamma = -0.2, -0.06, 0.6, 0.2$, respectively. The overall 
probability of null hypothesis ($\gamma = 0$) is $\bar{P} = 0.27$, therefore 
the probability of $\gamma < 0$, $1-\bar{P} = 0.73$. Light grey and cyan 
curves are theoretical calculation including the uncertainty on $\Omega_{de}$, 
respectively for $\gamma = \pm 0.06, \pm 0.2$. In ({\bf b}), the pink curve 
presents simulated distribution for $\gamma = 0$ - a cosmological constant. 
For this dataset $1 - \bar{P} = 0.75$. The dark grey straight 
line is the theoretical expectation for ${\mathcal A}(z)$ when dark energy is a 
cosmological constant. Definition of other curves are the same as ({\bf a}). 
\label{fig:sncp}}
\end{figure}

Two examples of reconstruction of simulated data are shown in 
Fig.\ref{fig:theorycurve}. It is evident that despite deformation of the 
reconstructed curves due to numerical errors and uncertainties, the difference 
between convexity of curves for positive and negative $\gamma$ is mostly 
preserved and can be used visually or by using a slope detection algorithm to 
find the sign of $\gamma$. The simulated data is however much more uniform 
than presently available data, see Fig.\ref{fig:sncp}. Although for mid-range 
redshifts the data follow a curve similar to what is expected from FLRW 
cosmologies, artifacts appear close to boundaries and at high redshifts where 
the quality of data is worse. Moreover, visual inspections or slope detection 
lack a quantitative estimation of uncertainty of measured sign for 
$\gamma$. Another complexity of this cosmological sign detection problem is 
that not only the observable ${\mathcal A}(z)$ is noisy, it also varies. In 
signal processing, in most practically interesting cases the signal is 
constant but noisy. Therefore, usual binomial estimation of the probability 
or optimization of detection~\cite{signopt} are not applicable. 

Here we take another strategy, specially suitable for this cosmological sign 
detection task. The null hypothesis for dark energy is $\gamma = 0$. Assuming a 
Gaussian distribution for uncertainty of reconstructed ${\mathcal A}(z)$ from 
data and from simulated data for $\gamma = 0$ model, for 
each data-point we calculate the probability that the data-point belongs to 
the null hypothesis. To include the uncertainty of data, we integrate the 
uncertainty distribution from $-\sigma$ to $+\sigma$ around the mean value:
\be
P_i = \frac{1}{\sqrt{2\pi (\sigma_{0i}^2 + \sigma_{i}^2)}} \int_{{\mathcal A}_i-\sigma_i}^{{\mathcal A}_i+\sigma_i} dx e^{-\frac {(x - {\mathcal A}_{0i})^2}{2 (\sigma_{0i}^2 + \sigma_{i}^2)}} 
\label{probzero}
\ee
where ${\mathcal A}_i$ and $\sigma_i$ are from $i^{th}$ data-point, and 
${\mathcal A}^{0i}$ and $\sigma_{0i}$ from simulated null hypothesis model at 
the same redshift. Averaging over $P_i$ gives $\bar {P}$, an overall 
probability that the dataset corresponds to the null hypothesis. As 
$\gamma = 0$ is the limit case for $\gamma > 0$, $\bar {P}$ is also the 
maximum probability of $\gamma > 0$.

We have applied this sign detection algorithm to two supernova datasets: 
published data from Supernova Legacy Survey (SNLS)~\cite{snls} and low 
redshift supernovae ($z < 0.45$) of gold sample of re-analyzed supernovae 
data by Riess \etal 2004~\cite{sncp}. The reason for using only the low 
redshift subset of the latter compilation is that the scatter 
and uncertainty of the peak magnitude at higher redshifts is too large, and 
with numerical methods used in this work, it is not possible to recover a 
reasonably stable and smooth distribution for ${\mathcal A}(z)$. 

Fig.\ref{fig:sncp} shows ${\mathcal A}(z)$ obtained from these data. To 
estimate the effect of the 
reconstruction, we compare ${\mathcal A}(z)$ from data with simulated data as 
described in the caption of Fig.\ref{fig:sncp}. Simulated standard sources are 
at the same redshifts as in the datasets to make simulated samples as similar 
to data as possible.

\begin{figure}[ht]
\begin{tabular}{cc}
{\bf a} \epsfig{figure=\filepath/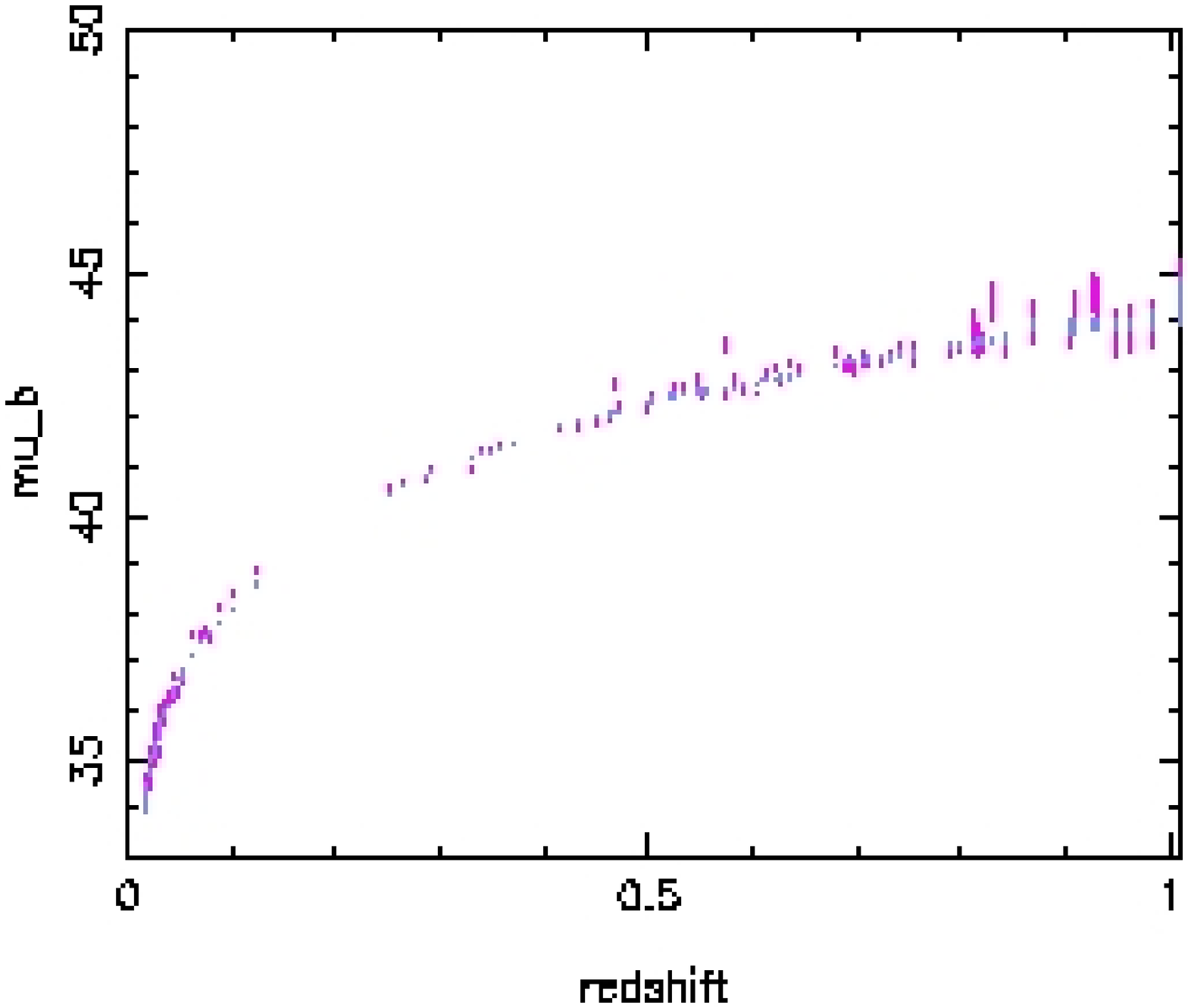,width=6.5cm} &
{\bf b} \epsfig{figure=\filepath/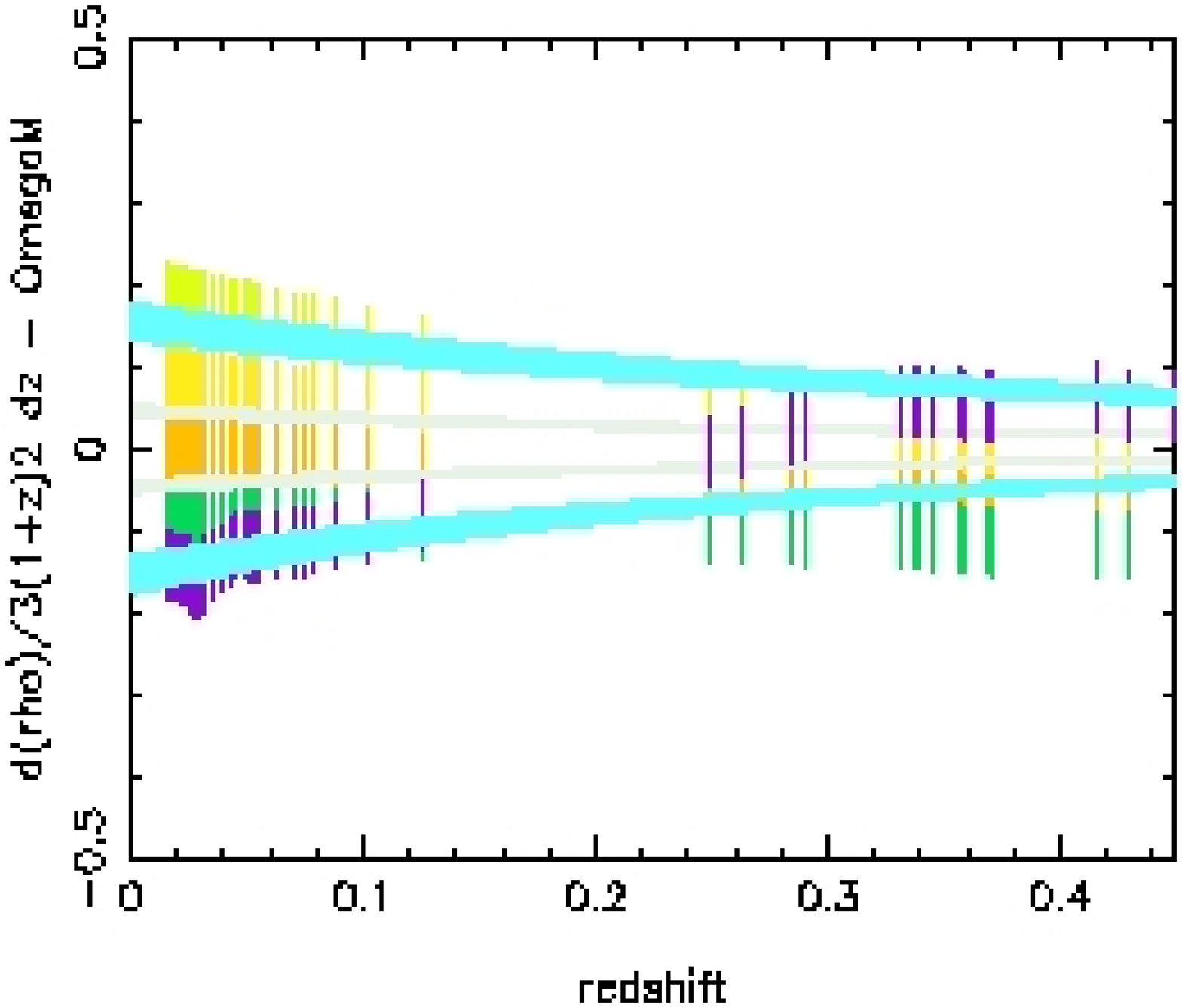,width=6.5cm}
\end{tabular}
\caption{{\bf a}: $\mu_b$ distribution of SNLS supernovae, data (magenta), 
smoothed distribution by the method explained in the Appendix (blue). Although 
this distribution look quite smooth, even small sudden variations can lead to 
large variations in derivatives. {\bf b}: ${\mathcal A}(z)$ for SNLS supernovae 
with $z < 0.45$. definition of curves is the same as in Fig.\ref{fig:sncp}. 
For this dataset $1-\bar{P} = 0.93$ when cosmological parameters are the same 
as Fig.\ref{fig:sncp}. If $\Omega_{de} = 0.73$ is used, $1-\bar{P} = 0.96$ for 
the same data.
\label{fig:snlslow}}
\end{figure}

In both datasets the probability of $\gamma \lesssim 0$ or 
equivalently $w \lesssim -1$ is larger than $70\%$. The SNLS data is 
consistent with a $\gamma$ as small as $\sim -0.2$\footnote{As the main 
purpose of this paper is determination of the sign of $\gamma$, we don't 
perform any fitting to obtain its values. Estimation of values here are from 
plots.}. There are however significant deviations from a smooth distribution 
for $z \lesssim 0.1$ and $z \gtrsim 0.5$. We attribute them to relatively large 
scatter of the data at these redshifts that makes reconstruction instable, 
see Fig.\ref{fig:snlslow}-{\bf a}. 

To see whether the large negative $\gamma$ concluded from the SNLS data is due 
to the data scattering and/or reconstruction algorithm artifacts, we also apply 
the same formalism to a subset of these data with $z < 0.45$. The result is 
shown in Fig.\ref{fig:snlslow}-{\bf b} along with simulations in the same 
redshift range. The {\it bump} at very low redshift in Fig.\ref{fig:sncp} does 
not exist in this plot, and therefore we conclude that it is an artifact of 
numerical calculation. Although ${\mathcal A}(z)$ distribution in this data 
set is also convex and the probability of $\gamma < 0$ is $ > 90\%$, it does 
not have the same slope as any of models. More specifically, it seems 
that low and high redshift sections of the curve correspond to different values 
of $\gamma$. For $z \lesssim 0.15$, ${\mathcal A}(z)$ is close to theoretical 
and simulated data with $\gamma = -0.2$. For $z \gtrsim 0.25$, 
${\mathcal A}(z)$ approaches the values for larger and even positive $\gamma$. 
Such behaviour does not appear in Fig.\ref{fig:sncp}. Giving the fact that the 
number of available data points with $z \gtrsim 0.25$ in this subset is small 
the most plausible explanation is simply numerical artifacts. Alternative 
explanations are evolution of $\gamma$ with redshift and an under-estimation 
of $\Omega_m$, see Eq.(\ref{densderiv}). If the latter case is ture, the value 
of $\gamma$ must be even smaller than $-0.2$. With a data gap in 
$0.15 \lesssim z \lesssim 0.25$ interval and a small total number of entries 
in this data set- only 58 supernovae - it is not possible to make any definite 
conclusion about the behaviour of this data. We should also mention that 
nearby SNe with $z < 0.25$, intermediate SNe $0.25 < z < 0.4$, and high 
redshifts ones $ z > 0.4$ in SNLS are not completely treated in the same 
way~\cite{snls}. It is therefore possible that some of the observed features 
explained here are purely artifacts of the analysis of raw observations.

Fig.\ref{fig:sncp}-{\bf b} shows ${\mathcal A}(z)$ determined from gold sample 
supernovae recompiled by Riess, \etal~\cite{sncp}. This dataset is consistent 
with $\gamma < 0$ with a probability $\sim 75\%$ at 1-sigma and $\sim 66\%$ at 
2-sigma. This plot also shows that the value of $\gamma$ estimated from this 
data is $\sim -0.06$, larger than estimation from SNLS data.

The reason for the difference between estimated values for $\gamma$ from SNLS 
and Riess, \etal compilation is not clear because the value of $w$ 
reported in Ref.~\cite{sncp} and Ref~\cite{snls} are consistent. Nonetheless, 
the estimated values here are in the range reported by other 
works~\cite{wmapeyear}~\cite{cmagicsn}. The difference between results of two 
datasets is 
most probably related to their different scatter and uncertainty, and the fact 
that low and high redshift data are not treated in the same way. We should 
also mention that recent claims about contamination of supernova type Ic and 
the effect of asymmetric explosion in the lightcurve of supernova type 
Ia~\cite{iaic}, and possible differences between low and high redshift 
supernovae~\cite{snlowhigh} can not explain the difference between the results 
of these datasets. These phenomena affect both surveys in the same way. 
Despite these discrepancies and uncertainties, both datasets are in good 
agreement about negative sign of $\gamma$.

The results of present study show various shortcomings in both datasets 
used here. Our first remark is the large gap in redshift distribution of 
observed supernovae in redshift range $0.1 \lesssim z \lesssim 0.3$. Both 
datasets have less than 6 supernovae in this range and nothing in 
$0.15 \lesssim z \lesssim 0.25$. This is not an important issue for finding 
redshift distribution of $D_l$, but redshift gap becomes very important when 
derivatives of $D_l$ are calculated. The lack or rareness of supernovae data in 
this redshift range is related partly to the history of star formation 
in galaxies~\cite{sfhist}, and partly to optimization of surveys~\cite{snzdist} 
for detection of very low or high redshift supernovae which decreases the 
probability of detection of mid-range SNe. A systematic survey of galaxies in 
this range is therefore necessary to fill the present gap. Sloan Supernova 
Survey~\cite{sdsssn} is optimized to detect SNe in 
$0.1 \lesssim z \lesssim 0.35$ and should provide the missing data in near 
future. 

Our second 
observation is a large scatter in both datasets around redshift $\sim 0.55$, 
see Fig.\ref{fig:snlslow}-{\bf a}. This leads to a large scatter in the 
numerical determination of derivatives in (\ref {rhoderdl}) and makes the 
results unusable, see Fig.\ref{fig:sncp}-{\bf a}. In future observations 
the reason of this large scatter should be understood, and if possible 
reduced. From theoretical calculation and simulations shown in 
Fig.\ref{fig:sncp} one can also conclude that with present uncertainties of 
cosmological parameters, the most important redshift range for determining 
the equation of state of dark energy is $z \lesssim 0.8$. Although the 
technical challenge, understanding of physics of supernovae~\cite{snlowhigh} 
and their evolution, and applications for other astronomical ends make the 
search of supernovae at larger distances interesting, they will not be in 
much use for determining the equation of state of dark energy, at least not 
at lowest level which is the determination of redshift independent component 
of $w$.

On the other hand, improvements in numerical techniques and algorithms would 
lead to better measurements. One of the possibilities in this direction 
is the application of an adaptive smoothing algorithm with variable degrees of 
smoothing depending on the amount of scattering in the data. More sophisticated 
smoothing algorithm also has been proposed~\cite{snsmooth}. We postpone the 
application these more advanced methods to future, when a larger data set 
becomes available.

In summary, we have proposed a nonparametric formalism to investigate the 
sign of $\gamma$ in the evolution equation of dark energy. When better quality  
data become available fitting can be added to to this method to find the value 
of $\gamma$ and not just its sign. The advantage of this method with respect to 
multi-parameter fitting is that the dependence on cosmological parameters 
is explicit, and therefore it is easier to assess the effect of their 
uncertainties on the measurements. This method is specially suitable for 
applying to supernova data where the standard observable - the peak magnitude - 
can be directly related to cosmological distance, and thereby to cosmological 
variation of total density. It can also be applied to data from galaxy 
clustering surveys which permit to determine the variation of average density 
of the Universe with redshift, but not to integrated observables such as CMB 
anisotropy. By applying this method to two of largest publicly available 
supernovae data sets we showed that they are consistent with a $w < -1$. 
Present data is not however enough precise to permit the estimation of $|w|$ 
with good certainty. With on-going projects such as SNLS, Supernova 
Cosmology Project~\cite {realsncp}, and SDSS SNe survey, and future projects 
such as SNAP and DUNE, enough precise datasets should be available 
in the near future.

{\bf Appendix:} To calculate ${\mathcal A}(z)$ in (\ref{densderiv}) for 
standard candles we must determine the luminosity distance. It is related to 
the magnitude of the standard candle: $D_l/D_0 = 10^{\mu_b / 5}$, where $D_0$ 
is the distance for which the common luminosity of standard sources is 
determined from theoretical models or observations. The standard magnitude at 
a given distance depends on $H_0$ and a correction must be added if a 
different $H_0$ is used in the calculation of $\mu_b$~\cite{snls}. For 
simulating the data $D_l$ is calculated from (\ref{dlh}) and the relation 
above is used to determine the corresponding $\mu_b$ to which we add a random 
uncertainty with a standard deviation of 3\%. From this point the same 
procedure is applied to both simulated and real data to determine 
${\mathcal A}(z)$.

Expression (\ref{rhoderdl}) for ${\mathcal B}(z)$ contains first and second 
derivatives of $D_l$ which must be calculated 
numerically from data. It is however well known that direct 
determination of derivatives leads to large and unacceptable deviation from 
exact values. One of the most popular alternative methods to the direct 
calculation is fitting of a polynomial around each data point and then 
calculating an analytical derivative using the polynomial approximation in 
place of the data. We use this approach to determine derivatives of $\mu_b$ 
and $D_l$. In addition, before applying this approach, we smooth the 
distribution of magnitudes using again the same polynomial fitting algorithm. 

In FLRW cosmologies the redshift evolution of the luminosity distance is 
very smooth. Therefore, a second order polynomial for smoothing is adequate.
Fitting is based on a right-left symmetric, least $\chi^2$ algorithm, and 
we have implemented {\it lfit} function of Numerical Recipes~\cite{numrec} 
for this purpose. By trial and error we find that 19 data-point fitting gives 
the most optimal results regarding the number of available data points and 
their scatter. Close to boundaries however less data point for fitting is 
available in one side of each point, and therefore the fitting is less precise. 
The artifacts discussed above are mostly related to this imprecision of 
numerical calculation. In the present work no adaptive smoothing is applied. In 
addition to smoothed data and their derivatives, the function {\it lfit} 
calculates a covariant matrix for uncertainty of parameters (derivatives). We 
use diagonal elements as 1-sigma uncertainty of the smoothed data and its 
derivatives. The uncertainty of ${\mathcal A}(z)$ is calculated from the 
uncertainty of terms in (\ref{densderiv}) and (\ref{rhoderdl}) using error 
propagation relation: For $f(x_1, x_2, \ldots)$, $\sigma_f^2 = \sum_i 
\sigma^2_{x_i} (\partial f/\partial x_i)^2$. Smoothed terms and parameters 
in ${\mathcal A}(z)$ are considered as independent variables with their own 
uncertainty.

\end{document}